\documentclass[sigconf]{acmart}
\usepackage[ruled,vlined]{algorithm2e}
\usepackage{multirow}
\usepackage{bm}
\usepackage{soul}
\usepackage{caption}
\usepackage{subcaption}

\AtBeginDocument{%
  \providecommand\BibTeX{{%
    \normalfont B\kern-0.5em{\scshape i\kern-0.25em b}\kern-0.8em\TeX}}}

\setcopyright{acmlicensed}
\acmJournal{PACMCGIT}
\acmYear{2021} \acmVolume{4} \acmNumber{1} \acmArticle{} \acmMonth{5} \acmPrice{15.00}\acmDOI{10.1145/3451254}

\acmSubmissionID{2}

\definecolor{tmp}{rgb}{0.6,0.,0.6}
\definecolor{myBlack}{rgb}{0,0.,0}
\definecolor{myRed}{rgb}{1,0.,0.}
\definecolor{kkBlue}{rgb}{0,0.0,0.9}
\definecolor{likeGreen}{RGB}{39, 191, 24}

\begin{document}


\title{Efficient Hyperparameter Optimization for\\ Physics-based Character Animation}

\author{Zeshi Yang}
\affiliation{%
  \institution{Simon Fraser University}
   \country{Canada}
}
\email{zeshiy@sfu.ca}

\author{Zhiqi Yin}
\affiliation{%
  \institution{Simon Fraser University}
   \country{Canada}}
\email{zhiqiy@sfu.ca}

\begin{abstract}
  Physics-based character animation has seen significant advances in recent years with the adoption of Deep Reinforcement Learning (DRL). However, DRL-based learning methods are usually computationally expensive and their performance crucially depends on the choice of hyperparameters. Tuning hyperparameters for these methods often requires repetitive training of control policies, which is even more computationally prohibitive. In this work, we propose a novel Curriculum-based Multi-Fidelity Bayesian Optimization framework (CMFBO) for efficient hyperparameter optimization of DRL-based character control systems. Using curriculum-based task difficulty as fidelity criterion, our method improves searching efficiency by gradually pruning search space through evaluation on easier motor skill tasks. We evaluate our method on two physics-based character control tasks: character morphology optimization and hyperparameter tuning of DeepMimic. Our algorithm significantly outperforms state-of-the-art hyperparameter optimization methods applicable for physics-based character animation. In particular, we show that hyperparameters optimized through our algorithm result in at least 5x efficiency gain comparing to author-released settings in DeepMimic.
\end{abstract}




\maketitle


\section{INTRODUCTION}

Physics-based character animation has made significant progresses recently, especially with the application of Deep Reinforcement Learning (DRL) algorithms \cite{peng2018deepmimic,bergamin2019drecon,won2020scalable,park2019learning}. For example, DeepMimic-style neural network controllers are able to synthesize diverse and robust high-quality motor skills \cite{peng2018deepmimic}. Despite the demonstrated impressive performance, it is usually quite hard for a novice graduate student to reproduce the performance of such systems, without knowing the exact value of each hyperparameter involved. Therefore, more and more authors have released their code for better reproducibility. However, a single change of one hyperparameter may totally ruin the performance, and sometimes even the convergence, of such learning algorithms. How the original authors found the working set of hyperparameters remains as an art and mystery. Improving upon prior work, even with released code, thus remains challenging, as any modification to the training algorithm may require a new set of hyperparameters to work well.

Automatic hyperparameter optimization is thus in great need. Hyperparameters, in the narrow sense, refer to values that are used to control the learning process in machine learning algorithms. In contrast, regular parameters are derived or optimized during the training process. In this paper, we refer to all parameters external to a learning algorithm that need to be determined prior to the learning as hyperparameters. For example, morphology parameters of a virtual character in motor learning are also hyperparameters. To date, it is a common practice of the field to manually test and select hyperparameters for various character control algorithms. Behind the scenes, maybe simple grid search or random search algorithms are used for semi-automatically choosing hyperparameters. However, in physics-based character control, better search schemes are needed to handle two key challenges: First, the evaluation of new hyperparameters typically involves re-learning of the controllers from scratch, which is usually an expensive black-box function. Second, the number of hyperparameters can also be large and results in the curse of dimensionality.

Bayesian optimization (BO) is a promising candidate for hyperparameter optimizations for physics-based character animation problems. BO is a sequential design strategy for global optimization of expensive-to-evaluate black-box functions that do not assume any functional forms \cite{jones1998efficient,srinivas2010gaussian,kandasamy2017multi}. Traditional BO only evaluates the expensive black-box objective functions themselves \cite{jones1998efficient,srinivas2010gaussian}. We refer to such algorithms as Single Fidelity BO (SFBO) in this paper. For many physics-based character animation applications, however, SFBO is inefficient due to the extremely high cost of function evaluations that rely on physics-based character simulation and motor learning. Multi-fidelity BO (MFBO) accelerates the optimization by using cheap approximations of the objective functions in early optimization stages \cite{kandasamy2017multi,klein2017fast,song2019general,swersky2013multi}. Such MFBO algorithms typically employ fewer training iterations on smaller datasets as low-fidelity cheap approximations to the original objective functions, and work well for supervised learning tasks.

For physics-based character animation, however, existing MFBO methods do not work well. We analyze in section \ref{sec::mffunction} and \ref{sec::pafo} that training iterations are not a good fidelity criterion for physics-based character control, and multi-fidelity objective does not have desired properties across fidelity dimension required by existing MFBO methods like BOCA \cite{kandasamy2017multi} and FABOLAS \cite{klein2017fast}. Therefore, we propose a novel algorithm CMFBO: Curriculum-based Multi-Fidelity Bayesian Optimization for efficient hyperparameter optimization of DRL-based character control systems. CMFBO employs easier motor skill learning tasks as low-fidelity optimization objectives. Task difficulties are organized and scheduled by a curriculum \cite{bengio2009curriculum}. For example, a curriculum that gradually increases the training episode length can help the character perform longer and longer skills \cite{peng2018deepmimic,pytorchrl}; and a curriculum that gradually reduces the hand-of-God assistance forces can help the character learn to locomote \cite{yu2018learning}. Easier tasks are faster to evaluate enabling efficient search space pruning. Control policies learned at easier tasks could be transferred to harder tasks to further reduce evaluation cost.
Hyperparameters optimized by CMFBO may significantly outperform those tuned by experts and optimized by state-of-the-art methods. For instance, at least 5x efficiency gain is obtained with our optimized hyperparameters on DeepMimic \cite{peng2018deepmimic}, compared to author-released settings.

To summarize, we
\begin{itemize}
    \item introduce Bayesian Optimization into physics-based character animation for principled hyperparameter optimization; 
    \item propose an efficient algorithm CMFBO for hyperparameter optimization of challenging motor learning tasks based on deep reinforcement learning;
    \item systematically compare and evaluate SFBO, state-of-the-art MFBO, and CMFBO on two physics-based character animation problems: morphology optimization, and automatic tuning of DeepMimic training hyperparameters. 
\end{itemize}

\section{RELATED WORK}

\subsection{Physics-based Character Animation}
Synthesizing natural human motions is a long-standing challenge in computer graphics. We classify existing methods roughly into three categories. First, manually designed controllers, which usually employ finite state machines (FSMs) and heuristic feedback rules. Human insights and domain knowledge are usually involved to design and tune the parameters \cite{hodgins1995animating,yin2007simbicon,jain2009optimization,wang2009optimizing,coros2010generalized,de2010feature,coros2011locomotion,wang2012optimizing,geijtenbeek2013flexible,felis2016synthesis}. Second, model-based trajectory optimization, where equations of motions are enforced as optimization constraints. Designing of an often sophisticated optimization objective function is generally required along with weights tuning for each term \cite{mordatch2012discovery,tassa2012synthesis,wampler2014generalizing,hamalainen2015online}. Recently, model-free DRL-based control methods have been demonstrated to reproduce high-quality complex and robust motor skills \cite{peng2015dynamic,lillicrap2015continuous,peng2016terrain,peng2017deeploco,heess2017emergence,peng2018deepmimic,peng2018sfv,yu2018learning,bergamin2019drecon,park2019learning,won2019learning,won2020scalable,luo2020carl,merel2020catch}. DRL-based methods require designing and tuning of a reward function, as well as related hyperparameters of DRL algorithms themselves. In this paper, We focus on hyperparameter optimization for recent DRL-based methods, although our framework is general enough to be applicable to other categories of methods as well.

\subsection{Hyperparameter Optimization}

\subsubsection{Parameter Optimization in Computer Graphics}
Parameter tuning and optimization is a common task in computer graphics, such as weights for SIMBICON-type feedback controllers \cite{yin2007simbicon,wang2009optimizing}. Multiple derivative-free optimization algorithms, such as Covariance Matrix Adaptation (CMA) \cite{hansen2006cma,wang2009optimizing} and Bayesian optimization, have been used to optimize such parameters where the objective functions are black-box functions and no derivative information is available. Some methods further combine BO with user evaluations to achieve desired visual effects, such as bidirectional reflectance distribution function (BRDF) design \cite{brochu2007preference} and smoke animation tuning \cite{brochu2010bayesian}. More recently, BO is applied to low-dimensional subproblems of various graphics applications \cite{koyama2017sequential,koyama2020sequential}, such as color enhancement, and human body geometric modelling \cite{loper2015smpl}.

We also adopt a Bayesian optimization framework, but for hyperparameter optimization. To the best of our knowledge, the only previous work on hyperparameter optimization in graphics is \cite{tseng2019hyperparameter}, where parameters of image processing hardware, such as threshold values in the denoising module, are optimized with a non-Bayesian approach. Hereafter we will focus on hyperparameter optimization literature from machine learning and robotics.

\subsubsection{Hyperparameter Optimization for DRL}
State-of-the-art motor learning methods are usually based on DRL. For example, DeepMimic \cite{peng2018deepmimic} can produce robust and diverse high-quality controllers by imitating motion captured reference motions. However, its performance crucially depends on the choice of its hyperparameters. Minor changes to one single hyperparameter may result in slow convergence or even training failure. But so far, no publications on physics-based character animation have explicitly discussed the issue of hyperparameter tuning or optimization. In machine learning, hyperparameter optimization for simple benchmark systems, such as a cartpole model in OpenAI gym \cite{greg2016gym}, with standard DRL algorithms, such as A2C \cite{mnih2016asynchronous}, has been investigated \cite{nguyen2019knowing,nguyen2020bayesian}. However, in physics-based character animation, we study the control of high dimensional character models, which is a much more expensive and challenging task requiring specialized DRL algorithms such as DeepMimic to achieve natural-looking skills.

\subsubsection{Morphology Design}
Morphology design and optimization is a problem in both computer animation and robotics \cite{park1994concurrent,sims1994evolving,pil1996integrated,lipson2000automatic,paul2001road,bongard2011morphological,villarreal2012robust,geijtenbeek2013flexible,agrawal2014diverse,spielberg2017functional,ha2017joint,wang2018nervenet,won2019learning,liao2019data,huang2020one,hu2020neural,Ma2021SpacetimeBounds}. Traditional model-based methods require accurate dynamic models, and only work for specific types of control algorithms such as trajectory optimization \cite{spielberg2017functional,ha2017joint}. Some recent morphology design methods work with DRL-based control \cite{schaff2019jointly,ha2019reinforcement,luck2020data}. These methods guide optimization in morphology space based on gradient estimation \cite{schaff2019jointly, ha2019reinforcement}, or performance prediction of unseen morphology through a learned value network \cite{luck2020data}. We note that none of these methods address the issue of expensive evaluation per design. With effective search space pruning through cheap evaluation on easier tasks, our method significantly outperforms them in terms of sample efficiency. 

\subsection{Bayesian Optimization}

Bayesian Optimization is a class of methods for expensive black-box function optimization. Since no gradient information is available, functions are optimized purely through evaluations. A Bayesian statistical model, usually a Gaussian Process \cite{rasmussen2003gaussian}, is maintained to estimate value of the objective function along with the uncertainty. An \textit{acquisition function} is then repeatedly applied to query and evaluate the most promising and informative region based on existing estimation. 

With the advancement of deep learning, there is a great appeal for automatic hyperparameter optimization of deep learning models. BO demonstrates its potential on such tasks for its promising sample efficiency \cite{snoek2012practical,snoek2015scalable,nguyen2019knowing}. 
Recently many MFBO methods are proposed to further improve efficiency of hyperparameter optimization for supervised learning tasks \cite{swersky2013multi,swersky2014freeze,kandasamy2016gaussian,kandasamy2017multi,klein2017fast,takeno2019multi,song2019general} mainly by utilizing cheap approximations of the objective, such as neural network validation loss on smaller datasets with fewer training iterations.

However, most of these methods cannot scale well to motion control tasks. Due to stochasticity and complexity of DRL training, existing fidelity criterion such as the number of training iterations mislead the optimization a lot. Low-fidelity evaluations through early stopping usually provide false estimation on the relative performance of hyperparameters. 
Besides the fidelity criterion issue, existing MFBO acquisition functions
have strong assumptions on the shape of multi-fidelity objective function along the fidelity dimension. For example, BOCA \cite{kandasamy2017multi} assumes a flat fidelity dimension while FABOLAS \cite{klein2017fast} assumes it to satisfy quadratic form. These simplified assumptions are no longer valid in the complex DRL settings. We detail the analysis and how we resolve the problems with our proposed curriculum-based fidelity and progressive acquisition function in section \ref{sec::mffunction} and \ref{sec::pafo}.


\subsection{Curriculum Learning}
Curriculum learning (CL) is a learning method where task difficulty gradually increases during training\cite{bengio2009curriculum}. Easier tasks are cheaper to accomplish and serve as good initial solutions for more difficult tasks. CL has been widely used in character animation to improve the performance and efficiency of motor learning tasks \cite{van1995guided,yin2008continuation,wu2010terrain,karpathy2012curriculum}. \cite{peng2018deepmimic,pytorchrl} employ a time-based curriculum with increasing episode length to help the character perform longer and longer skills.
\cite{yu2018learning} applies hand-of-God assistance forces at robot torso and decreases the force gradually during training to encourage the emergence of natural gaits. More recently, \cite{xie2020allsteps} uses CL to demonstrate successful training of stepping-stone locomotion controllers.

\section{METHOD}
We formulate hyperparameter optimization in a physics-based animation system as a black-box function optimization problem
\begin{equation}
    \mathop{\arg\max}\limits_{x \in \mathcal{A}}f(x)
\end{equation}
where $\mathcal{A}$ is the hyperparameter space. $f(\cdot): \mathcal{A} \to \mathcal{R}$ is the objective function, i.e. performance of control policy given hyperparameter $x$. The evaluation of $f(\cdot)$ involves DRL training, thus is costly, noisy and cannot be computed in closed forms.

\begin{figure}[]
 \centering
 \includegraphics[width=\linewidth]{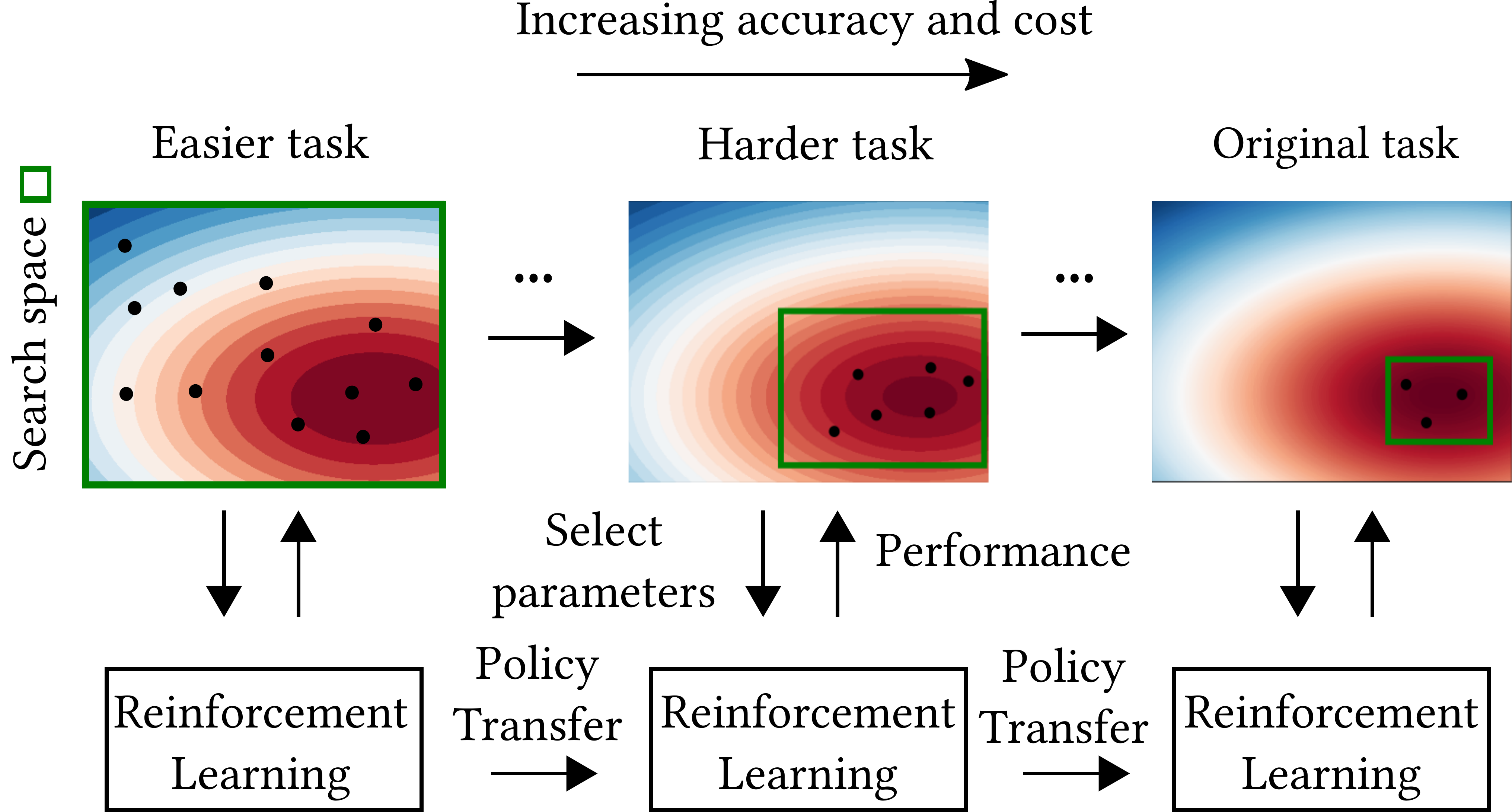}
 \caption{Conceptual illustration of CMFBO}
 \label{fig:pipeline}
\end{figure}

Figure \ref{fig:pipeline} illustrates a conceptual overview of our approach based on MFBO. We first define our multi-fidelity objective function based on a curriculum (section \ref{sec::mffunction}). During optimization, we use lower-fidelity cheap approximations to locate the promising region of hyperparameters through the proposed progressive acquisition function (section \ref{sec::pafo}). Policies learned at easier tasks are transferred to difficult tasks for more efficient evaluation (section \ref{sec::policytransfer}).

\subsection{Background}
\label{sec::background}

\begin{figure*}[]
  \centering
  \includegraphics[width=\linewidth]{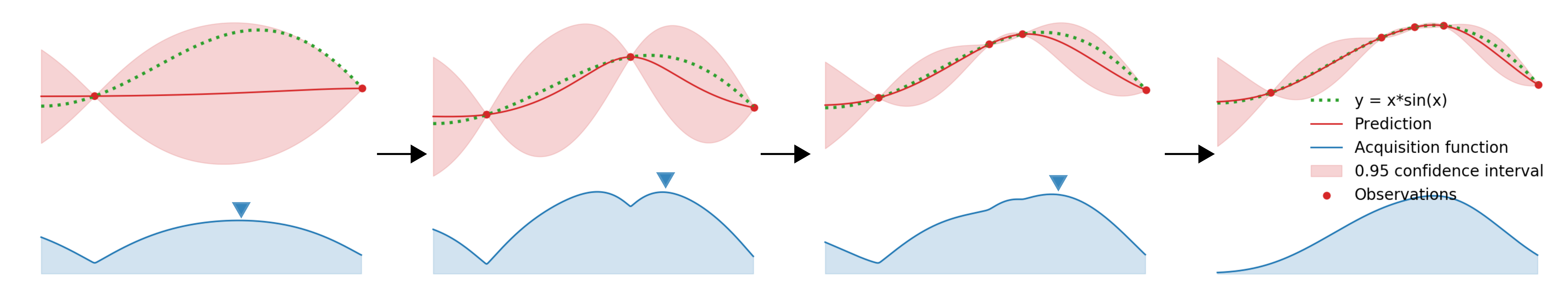}
  \caption{An example of applying BO to a one-dimensional function. Green dotted curve represents the unknown black-box function. Red points represent the queried points. Red curve along with the rosy shaded region indicate the predicted mean and $95\%$ confidence interval. Blue curve is the acquisition function (It is scaled for the convenience of visualization). Blue downside arrows indicate the maximum of the acquisition function which is the point to be queried next.}
  \label{fig:BO}
\end{figure*}

In this section we briefly review the technique of general Bayesian optimization involved in our approach.
Given a black-box function $f(x)$, BO finds its maxima through repeated function evaluations. Since evaluating $f(x)$ could be expensive, BO is designed to minimize the number of function evaluations by querying the most promising and informative points. Given a set of current observations $D_{t} = \{(x_{i}, y_{i})\}_{i=1}^{t}$, where $y_{i}$ is a noisy measurement of $f(x_{i})$, the \textit{acquisition function} $a(x, D_{t}): \mathcal{A} \to \mathbb{R}$ quantifies the utility of an arbitrary point. Maximizing the acquisition function will give us the point most worth trying next.


The acquisition function is designed for finding candidate points with both large values and rich information. Querying a point close to existing ones in $D_{t}$ is less informative. Gaussian Process Upper Confidence Bound (GP-UCB) \cite{srinivas2010gaussian} is a popular acquisition function defined as:
\begin{equation}
    a(x, D_{t}) = \mu_{t}(x) + \beta^{\frac{1}{2}} \sigma_{t}(x)
\end{equation}
where $\mu_{t}(x)$ and $\sigma_{t}(x)$ are approximated posterior mean value and standard deviation of $f(x)$ respectively.
$\mu_{t}(\cdot)$ favors candidates which are likely to have large values. $\sigma_{t}(\cdot)$ encourages querying informative points with high uncertainty. $\beta$ enables a trade-off between exploitation and exploration.

Closed-form estimations for $\mu_{t}(\cdot)$ and $\sigma_{t}(\cdot)$ are available through a Gaussian Process (GP) \cite{rasmussen2003gaussian} surrogate model of the objective function trained on $D_{t}$.
A GP contains a prior mean function $m(x)$ and a kernel function $k(x,x')$. $m(x)$ encodes our prior belief of the objective function value. Kernel function $k(x,x')$ measures correlations between $f(x)$ and $f(x')$. Given $m(\cdot)$, $k(\cdot, \cdot)$ and existing observations $D_{t}$, $\mu_{t}$ and $\sigma_{t}$ could be computed as:
\begin{equation}
    \begin{gathered}
        \mu_{t}(x) = k(x,X) (K + \eta^{2} I)^{-1} Y \\
        \sigma_{t}^{2}(x) = k(x, x) + \eta^{2} - k(x,X)(K + \eta^{2}I)^{-1} k(X,x)
    \end{gathered}
\end{equation}
where $Y \in \mathbb{R}^{t}, Y_{i} = y_{i}$; $X \in \mathbb{R}^{t \times d}, X_{i} = x_{i} $; $K \in \mathbb{R}^{t \times t}, K_{i,j} = k(x_{i}, x_{j})$; $k(x,X) = (k(x,x_{1}),k(x,x_{2}),...k(x,x_{t}))$; $\eta$ is the standard deviation of the observation noise.

Many choices of kernel functions exist, including Square Exponential kernel and Matern kernel. In this work we adopt the Square Exponential kernel defined as:
\begin{equation}
    k_{SE}(x, x') = \sigma_{f}^{2}  exp(-\frac{1}{2}(x-x')^{T}\Lambda^{-1}(x-x'))
\end{equation}
$\Lambda = diag(\lambda_{1}, \lambda_{2},...\lambda{t})$, where $\lambda_{i}$ is the length scale of the $i$-th component of inputs. $f(\cdot)$ is flat across the $i$-th component of the input when $\lambda_{i}$ is large.

Figure \ref{fig:BO} shows a visual illustration of applying BO to a one-dimensional function. We refer interested readers to the second chapter of \cite{rasmussen2003gaussian} for more in-depth reviews.




\subsection{Curriculum-based Multi-Fidelity Functions}
\label{sec::mffunction}

The main challenge of hyperparameter optimization for physics-based character control tasks comes from the fact that performance evaluation function $f(x)$ given hyperparameter $x$ is highly computationally expensive. Even if BO is designed for black-box function optimization with minimum samples, the entire optimization procedure still requires a prohibitive amount of computational resources.

\begin{figure}
    \setlength{\abovedisplayskip}{1pt}
    \setlength{\belowdisplayskip}{1pt}
     \centering
     \begin{subfigure}[]{0.4\linewidth}
         \centering
         \includegraphics[width=\linewidth]{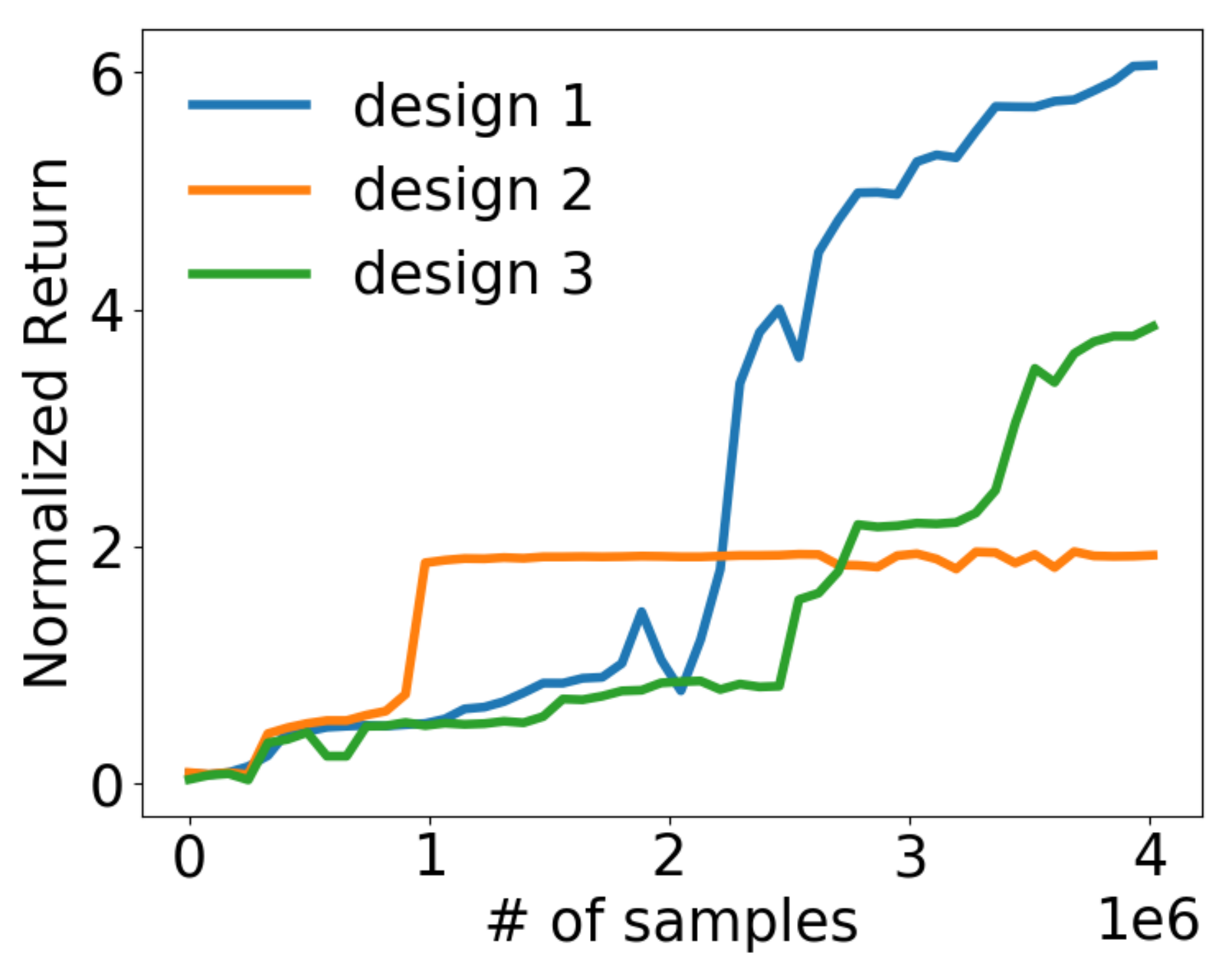}
         \caption{}
         \label{fig:bad_fidelity}
     \end{subfigure}
     \begin{subfigure}[]{0.4\linewidth}
         \centering
         \includegraphics[width=\linewidth]{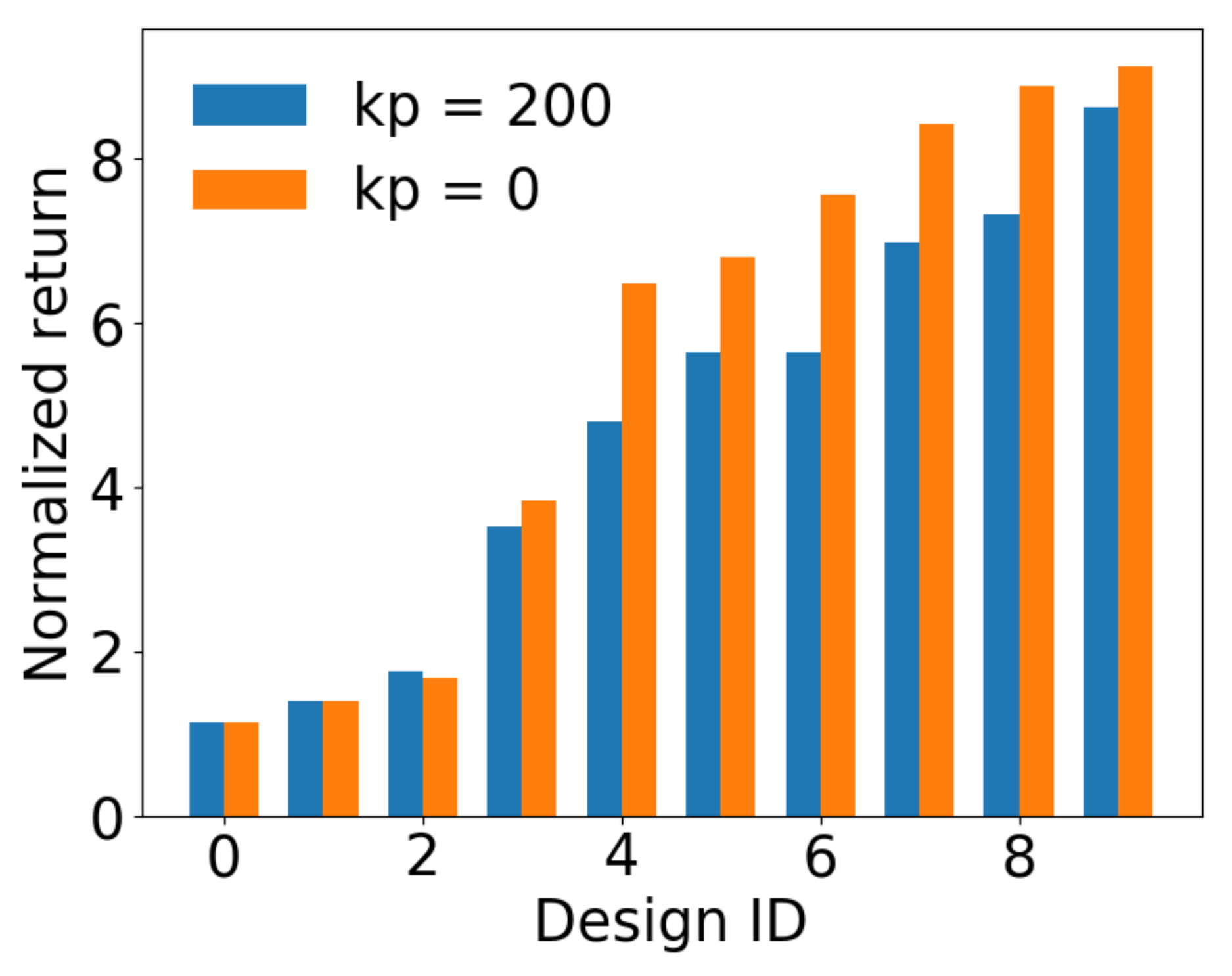}
         \caption{}
         \label{fig:good_fidelity}
     \end{subfigure}
         \caption{(a): Learning curves of locomotion controllers with different morphology designs. (b): Performance of ten morphology designs on two difficulty levels. Task difficulty is determined by the strength of hand-of-God assistance forces parameterized by proportional gain $\textit{kp}$ of the assistance stable PD controller.}
        \label{fig:fidelity}
\end{figure}

We propose to solve this problem with a multi-fidelity approach on top of Bayesian optimization, where low-fidelity cheap evaluations are utilized for efficient search space pruning. This procedure crucially depends on an accurate fidelity criterion being able to distinguish good hyperparameters from bad ones with lower-fidelity approximations. To the best of our knowledge, number of training iterations is the only fidelity criterion from existing literature applicable to our problem. Though number of training iterations shows its effectiveness for supervised learning tasks \cite{kandasamy2017multi,klein2017fast, snoek2012practical}, performance estimated through early stopping can be deceptive for challenging DRL character control tasks. Early stages for DRL training can be noisy \cite{haarnoja2018soft,fujimoto2018addressing}, and good hyperparameters do not always guarantee quick convergence \cite{nguyen2020bayesian}. We illustrate this problem in the context of finding optimal morphology hyperparameters for character locomotion. We show the performance of three morphology designs in figure \ref{fig:bad_fidelity}, where some good morphology designs show worse performance in early training stages.

To mitigate this problem, we propose the use of curriculum-based task difficulty as a new fidelity criterion tailored to physics-based character control settings. Task difficulty can be parameterized by a single continuous scalar variable $z$. Multi-fidelity objective function is therefore defined to be $f(x,z)$. Note that $f(x,z_{max})$ is the original objective function. $f(x,z)$ is more accurate and requires more computational resources as $z$ gets closer to $z_{max}$. 
Similar to single fidelity BO, we use GP to model multi-fidelity function $f(x,z)$. The kernel function generalizes to:
\begin{equation}
\label{equation:kernel}
    k((x,z), (x',z')) = k_{SE}(x,x')\ k_{SE}(z,z')
\end{equation}
The factorized kernel adopts the multiplication form based on the fact that: if $x$ and $z$ are close to $x'$ and $z'$ respectively, $f(x,z)$ correlates strongly with $f(x',z')$.

We validate the effectiveness of using task difficulty as fidelity criterion through an example shown in figure \ref{fig:good_fidelity}, where ten morphology hyperparameters are evaluated on two different levels of task difficulty parameterized by the strength of virtual hand-of-God assistance forces. With fidelity criterion defined with task difficulty, relative performance of hyperparameters across different fidelity levels are well preserved. We detail our specific curriculum settings for different tasks in section \ref{sec::results}.

\subsection{Progressive Acquisition Function}
\label{sec::pafo}

Given a properly designed multi-fidelity objective function, an effective acquisition function is still in need to achieve efficient hyperparameter optimization in physics-based character control tasks. In each iteration, multi-fidelity acquisition function $a((x,z), D_t)$ determines which point $x$ in hyperparameter space is going to be evaluated next, and which fidelity level $z$ the evaluation is going to be performed on. An effective multi-fidelity acquisition function is expected to generate queries such that the original objective can be optimized with much fewer samples by inferring optimality information through low-fidelity cheap approximations. 


Existing MFBO methods like BOCA \cite{kandasamy2017multi} and FABOLAS \cite{klein2017fast} require a well-fitted surrogate model for accurate estimation of original objective value from lower-fidelity approximations. Mathematical assumptions on the shape of fidelity dimension are usually necessary to achieve this. For example, BOCA assumes the fidelity dimension is flat while FABOLAS assumes it to satisfy quadratic form. However, due to stochasticity and complexity of DRL training, performance difference of hyperparameters on different fidelity levels can be unpredictable, as shown in figure \ref{fig:good_fidelity}. A GP trained on such DRL performance data will learn a small length scale, which degrades the accuracy of high fidelity value prediction through a reasonable number of lower-fidelity samples.   

We attempt this problem with a different approach. Instead of relying on low fidelity samples to infer high-fidelity function value, we propose to use them simply for locating promising regions. The motivation is based on the key observation that though performance difference of hyperparameters on different fidelity levels can be unpredictable, optimal regions of objective function on different fidelity levels highly overlap, as shown in figure \ref{fig:good_fidelity}. Search space for a high-fidelity objective can therefore be greatly pruned by simply exploring around optimal regions estimated at lower fidelity levels. 

Specifically, we propose a novel progressive acquisition function, which progressively conducts optimization from low fidelity to high fidelity levels. At iteration $t$, target fidelity $z_t$ is chosen to be the least expensive while most informative one until we reach the highest. This is performed through an iterative procedure. Initializing $z_t$ with $z_{min}$, we repeatedly increase $z_t$ by a small step $\alpha$ and re-compute best candidate point $x_t$ for new $z_t$. The procedure terminates when our model shows insufficient knowledge at $(x_t, z_t)$ with uncertainty $\sigma_{t}(x_t, z_t)$ exceeding threshold $\epsilon$. In our implementation, we choose $\alpha$ to be equal to the length scale of the fidelity kernel $l_z$, since function values from GP surrogate model strongly correlate along fidelity dimension within $l_z$ proximity. $z_t$ should be increased only when we have sufficient knowledge at $(x_t,z_t)$. $\epsilon$ can therefore be set conservatively to a small value negligible comparing to the objective function value range in promising regions. In our experiments we set $\epsilon$ to $0.02$.

Given a specified fidelity level $z_t$, the best candidate point $x_t$ is chosen by optimizing a variant of the upper confidence bound (UCB) utility function defined as: 
\begin{equation}
    a((x,z_t),D_{t}) = \mu_{t}(x,z_t) + \beta^{\frac{1}{2}}\sigma_{t}(x,z_{max})
    \label{equ:acquistion}
\end{equation}
where $\mu_{t}(x,z_t)$ favors high-performing regions and $\sigma_{t}(x,z_{max})$ favors uncertain regions of the original objective. We set $\beta$ to $0.2 d\log(2t)$ following suggestions from \cite{kandasamy2017multi,srinivas2010gaussian}, where $d$ is the dimension of the hyperparameters and $t$ is the iteration index.
The optimization starts from $z_{t} = z_{min}$ to find initial promising candidates across the hyperparameter space. When $z_t > z_{min}$, optimization of $a((x,z_t),D_{t})$ is restricted to be close to known high-utility regions obtained from lower-fidelity levels, which enables efficient search space pruning. This can be achieved by simply applying a gradient-based local optimization method like L-BFGS \cite{liu1989limited} on $a((x,z_t),D_{t})$, with solutions of $\mathop{\arg\max}a(\cdot)$ at lower-fidelity levels serving as initial guesses. The optimization of the acquisition function does not have to be perfect, since we only care about promising points instead of optimal ones. Our algorithm works well with our default choices for $\alpha,\beta$ and $\epsilon$ without manual tuning. We summarize this procedure in algorithm \ref{algorithm::PAF} for ease of re-implementation.

\subsection{Policy Transfer}
\label{sec::policytransfer}
Progressive acquisition function enables the use of transfer learning to speed up evaluation on most $(x_{t},z_{t})$ pairs where $z_t > z_{min}$. This comes from the fact that progressive acquisition function queries $z_t$ progressively, and only looks for $x_t$ within a common promising region. Therefore, as long as $z_t > z_{min}$, a previously queried pair $(x_i,z_i)$ with $i < t$ usually (if not always) exists such that $(x_{t},z_{t})$ is close to $(x_i,z_i)$. Pre-trained policy on $(x_i,z_i)$ can therefore serve as a warm start for training on $(x_{t},z_{t})$. 

In practice, $(x_i,z_i)$ can be acquired by traversing all previously visited $(x,z)$ pairs to find the closest one to $(x_{t},z_{t})$ measured by kernel function correlation $k((x_{t},z_{t}), (x_i,z_i))$. Theoretically, negative transfer may happen but we do not observe any in our experiments. We show in our experiments that the use of transfer learning saves a significant number of samples during CMFBO optimization.

Algorithm \ref{algo::CMFBO} summarizes the pipeline of CMFBO.

\begin{algorithm}
\caption{Progressive Acquisition Function }\label{algorithm::PAF}
\KwIn{Iteration index $t$, fidelity kernel length scale $l_{z}$, observation dataset $D_{t}$ and uncertainty threshold $\epsilon$ }
\KwOut{Best candidate pair $(x_{t}, z_{t})$}
$z_{t} \leftarrow z_{min}$
\\
$x_{t} \leftarrow \mathop{\arg\max}\limits_{x}a((x,z_{t}), D_{t})$ using L-BFGS with multiple random start points\\
\While{$z_{t} < z_{max}$ and $\sigma_{t}(x_{t},z_{t}) < \epsilon$}
{
1.$z_{t} \leftarrow \mathop{\min}(z_{t} + l_{z}, z_{max})$ \;
2.$x_{t} \leftarrow \mathop{\arg\max}\limits_{x}a((x,z_{t}), D_{t})$ using L-BFGS with $x_{t}$ as an initial solution \;}
Return $(x_{t}, z_{t})$
\end{algorithm}


\begin{algorithm}
\caption{Curriculum-based MFBO}\label{algorithm::CMFBO}
\KwIn{Multi-fidelity function $f(x,z)$ and the maximum allowed computational cost $N_{total}$}
\KwOut{Observed function maximum $y^{\star}$ and corresponding $x^{\star}$}
$N_{current} \leftarrow 0, D_{0} \leftarrow \varnothing$, the Gaussian Process is initialized randomly. \\
\For{$t = 1,2,...$}
{1.Choose $(x_{t},z_{t})$ by progressive acquisition function optimization (see algorithm \ref{algorithm::PAF})\;
2.Evaluate $f(x_{t},z_{t})$ to get performance $y_{t}$ and corresponding computational cost $N_{t}$\;
3.$D_{t} \gets D_{t-1} \cup {((x_{t},z_{t}), y_{t})}$ and update parameters of the Gaussian Process using $D_{t}$\;
4.$N_{current} \gets N_{current} + N_{t}$\;
5.\uIf{$N_{current} \ge N_{total}$}
{break;}
}
$t^{\star} \gets \mathop{\arg \max}\limits_{t}y_{t}$, $y^{\star} \gets y_{t^{\star}}$, $x^{\star} \gets x_{t^{\star}}$\;
Return $y^{\star}$ and $x^{\star}$
\label{algo::CMFBO}
\end{algorithm}
\section{EXPERIMENTS}
\label{sec::results}
We validate CMFBO on two physics-based character control tasks: morphology optimization and hyperparameter optimization of DeepMimic. 
For each task, we discuss its specific curriculum setting and objective function, followed by experiment results, comparisons and ablation studies.

In our implementation, we use Gaussian Process from GPy\cite{gpy2014} and L-BFGS \cite{liu1989limited} from Scipy \cite{virtanen2020scipy}.
All physics simulation is performed on PyBullet\cite{coumans2016pybullet}.
We report the performance statistics on a desktop with i7-7800x CPU (12 threads). Each experiment is run $3$ times independently with different random seeds.

\subsection{Morphology Optimization}
The locomotion capability of a creature is intimately coupled with its morphology. Elaborately designed morphology of simulated characters encourages the emergence of natural locomotion \cite{geijtenbeek2013flexible}.
In this experiment, we optimize character morphology for learning fast and low-energy locomotion. Our simulated character has $15$ torque-controlled revolute joints.
Morphology hyperparameter $x$ is defined to be a $12$-dimensional vector that scales the length and radius of character links within a limited range, as shown in table \ref{table::paras_morphology}. Link mass is uniformly scaled according to its volume. Sagittal symmetry is imposed on legs of the character. 

\subsubsection{Environment Setup}
\label{sec::morphology_env_setup}
We designed a DRL-based character control environment tailored to fast locomotion, based on humanoid environments in \cite{coumans2016pybullet}. 
Character state is defined to be orientation and linear velocities of the root link, along with angles and velocities for all joints. Action is normalized joint torques. We adopt a reward function specialized for locomotion control proposed in \cite{xie2020allsteps}, except that task specific and velocity penalty terms are removed to encourage faster locomotion. Readers can refer to \cite{xie2020allsteps} for more details of the reward function. 
We further adopt the symmetry loss
proposed in \cite{yu2018learning} to encourage symmetric gaits.

We use a feed-forward policy network with three fully connected layers, each with $128$ units using $\mathop{\mathrm{tanh}}$ activation. The critic network shares the same architecture with the policy. Policy is trained with Proximal Policy Optimization (PPO) \cite{schulman2017proximal}. We set discount factor $\gamma = 0.95$ and $\lambda = 0.95$ for both $TD(\lambda)$ and $GAE(\lambda)$ \cite{schulman2015trust}. Learning rates of both policy and critic network are set to $3\times10^{-4}$. In each training iteration, we sample $4096$ state-action tuples with $10$ paralleled environments. Batch size for policy update is set to $256$.

\subsubsection{Curriculum and Multi-fidelity Objective}
\label{sec::morphology_character}
We construct our multi-fidelity function based on a curriculum learning setting proposed in \cite{yu2018learning}, where a stable proportional-derivative (SPD) controller \cite{tan2011stable} is applied to the character root to provide hand-of-God balancing forces. Task difficulty is parameterized by stiffness $\textit{kp}$ and damping $\textit{kd}$ coefficients of SPD controller. High gain controllers make the task easier by providing large assistance forces. Given the normalized task difficulty scalar $z$ ranging from $0$ to $1$, we set $\textit{kp}(z) = 200 - 200 z$. $\textit{kd}$ is set to be equal to $\textit{kp}$ as proposed in \cite{yu2018learning}. The policy training starts at $\textit{kp}(0)$. \textit{kp} gradually decreases to $\textit{kp}(z)$ during training according to the curriculum schedule in \cite{yu2018learning}. We refer readers to section $4.2.2$ in \cite{yu2018learning} for more details. The multi-fidelity objective is the normalized return defined as:
\begin{equation}
    f(x, z) = \frac{\sum_{t=1}^{T}r(s_{t}, a_{t})}{T}
\end{equation}
where $r(s_{t},a_{t})$ is the reward function, and $T$ is the episode length fixed to $500$ simulation steps ($8.25$ seconds).

\begin{table}[]
\begin{tabular}{@{}cccc@{}}
\toprule
Length scale & Range                            & Radius scale & Range          \\ \midrule
Torso  & {[}0.3, 2{]}                     & Torso  & {[}0.5, 1.5{]} \\
Waist  & {[}0.3, 2{]}                     & Waist  & {[}0.5, 1.5{]} \\
Pelvis & {[}0.3, 2{]}                     & Pelvis & {[}0.5, 1.5{]} \\
Thigh  & \multicolumn{1}{l}{{[}0.3, 2{]}} & Thigh  & {[}0.5, 1.5{]} \\
Knee   & \multicolumn{1}{l}{{[}0.3, 2{]}} & Knee   & {[}0.5, 1.5{]} \\
Foot   & \multicolumn{1}{l}{{[}0.3, 2{]}} & Foot   & {[}0.5, 1.5{]} \\ \bottomrule
\end{tabular}
\caption{Morphology hyperparameters.}
\label{table::paras_morphology}
\end{table}

\begin{figure}
     \centering
     \begin{subfigure}[]{0.4\linewidth}
         \centering
         \includegraphics[width=\linewidth]{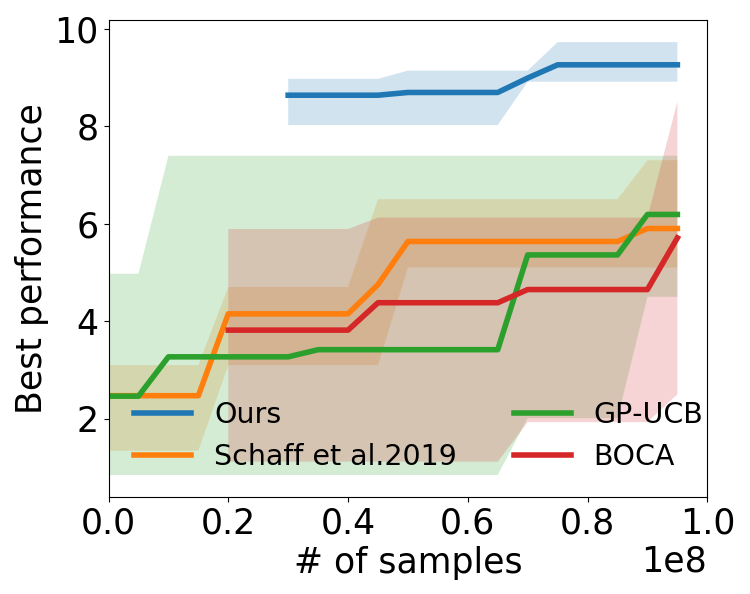}
         \caption{}
         \label{fig:lr_morphology_comparision}
     \end{subfigure}
     \begin{subfigure}[]{0.4\linewidth}
         \centering
         \includegraphics[width=\linewidth]{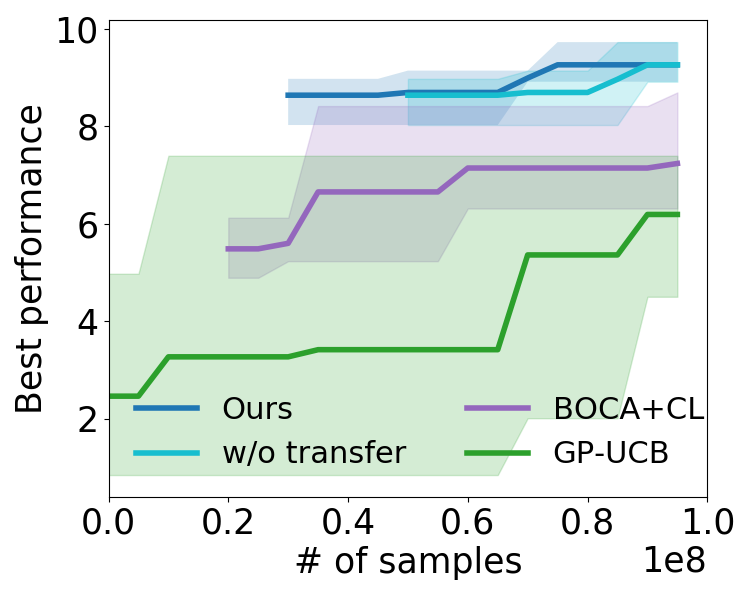}
         \caption{}
         \label{fig:lr_morphology_ablation}
     \end{subfigure}
         \caption{(a):Results on morphology optimization. We compare best performance over the number of training samples among CMFBO (Ours), GP-UCB, BOCA with number of iteration as fidelity and \cite{schaff2019jointly}. (b): Ablation study on morphology optimization. We further compare our method with BOCA with curriculum-based fidelity (BOCA+CL), and CMFBO without policy transfer (w/o transfer).}
        \label{fig:lr_morphology}
\end{figure}

\subsubsection{Comparison}
We compare our method with several baseline and state-of-the-art algorithms: a single fidelity BO with GP-UCB acquisition \cite{srinivas2010gaussian}, a multi-fidelity method BOCA \cite{kandasamy2017multi} and a recent DRL-based algorithm \cite{schaff2019jointly} (Schaff et al. 2019) specialized for morphology optimization. GP-UCB and \cite{schaff2019jointly} can be directly applied to the problem. For multi-fidelity method BOCA, We implement it following \cite{kandasamy2017multi} and construct the multi-fidelity function parameterized by both number of training iterations and our proposed curriculum-based task difficulty. Maximum allowed iteration number at fidelity $z$ for BOCA is set to $500 + 1500 z$ since most training finishes within $2000$ iterations.
Maximum allowed simulation steps $N_{total}$ is set to $10^{8}$. The optimization process takes about $18$ hours for all methods.

\begin{figure*}[]
  \centering
  \includegraphics[width=\linewidth]{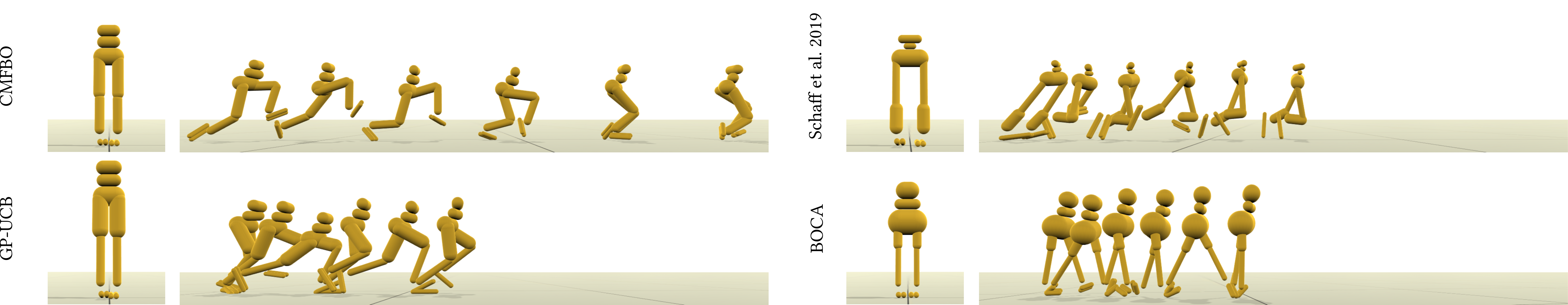}
  \caption{Visualization of morphology and gaits optimized by different methods.}
  \label{fig:comparison_morphology}
\end{figure*}

\subsubsection{Results}
We plot the best performance over number of total simulation steps in figure \ref{fig:lr_morphology_comparision} for various methods. Note that for multi-fidelity methods, best performance is only recorded when the original objective function is evaluated. Our method finds high-performing morphology hyperparameters much faster than other methods. BOCA performs worse than single fidelity method GP-UCB due to a misleading fidelity criterion defined as number of training iterations, as explained in section \ref{sec::mffunction}. Figure \ref{fig:comparison_morphology} visualizes the best morphology and the associated gaits for each method obtained from $3$ individual runs. Our method learns morphology resembling human and shows fast and stable gaits.
We encourage readers to see supplementary videos for more details. 
Note that though our method learns different morphology hyperparameters at different runs, proper values of essential components key to stable locomotion are consistently learned, like thigh and foot lengths.



Figure \ref{fig:lr_morphology_ablation} shows results of our ablation study, where we further compare our method against BOCA with our proposed curriculum-based fidelity (BOCA + CL), and our method without policy transfer (w/o transfer). 
With an effective fidelity criterion defined by curriculum-based task difficulty, BOCA shows a large performance gain. However, our method still outperforms BOCA by a large margin due to effective search space pruning through our progressive acquisition function. The use of transfer learning further improves efficiency of our method, with a saving of roughly $2\times10^{7}$ samples before a high-performing set of hyperparameters is confirmed on the original fidelity.

\subsection{DeepMimic Hyperparameter Optimization}
DeepMimic is a recent DRL-based framework for diverse motor skill learning. Many recent physics-based character animation systems are built on DeepMimic, such as \cite{park2019learning,bergamin2019drecon,won2019learning,won2020scalable,luo2020carl}. 
As far as we know, this is the first work towards hyperparameter optimization for high-dimensional physics-based character control systems. In our experiments, we focus on the optimization of DRL training hyperparameters listed in table \ref{table::paras_deepmimic}. The choices of these hyperparameters are critical for the learning performance where none of our randomly sampled $5$ hyperparameters lead to successful policy training. These hyperparameters correlate with each other implicitly and are hard to be manually tuned. Some recent works \cite{Ma2021SpacetimeBounds,won2019learning} therefore simply adopt the default DeepMimic hyperparameters directly into their systems with minor modifications. We evaluate our method on two motor skill learning tasks: walk and backflip. Our optimized hyperparameters improve DeepMimic learning efficiency by a large margin. Hyperparameters optimized for walking can be reused for other skills as well resulting in superior performance than original settings. 
\begin{table}[]
\begin{tabular}{ccc}
\hline
Parameters                      & Range            &Default\\ \hline
Batch size                       & {[}$32$,$1024${]}   &$256$ \\
\begin{tabular}[c]{@{}c@{}}Policy updates \\ per iteration\end{tabular}      & {[}$1$,$10${]}      &$1$\\
\begin{tabular}[c]{@{}c@{}}Learning rate \\ of actor network\end{tabular}  & {[}$2.5\times10^{-6}$,$2.5\times10^{-4}${]}       &$2.5\times10^{-6}$\\
\begin{tabular}[c]{@{}c@{}}Learning rate \\ of critic network\end{tabular} & {[}$1\times10^{-4}$,$1\times10^{-2}${]}       &$1\times10^{-2}$\\
Weight decay                    & {[}$5\times10^{-4}$,$5\times10^{-2}${]}       &$5\times10^{-4}$\\
PPO clip rate                   & {[}$0.02$, $0.2${]} &$0.2$\\
\begin{tabular}[c]{@{}c@{}}Maximum of \\ the gradient norm\end{tabular}    & {[}$1$, $100${]}    &$100$\\ \hline
\end{tabular}
\caption{Hyperparameters of DeepMimic}
\label{table::paras_deepmimic}
\end{table}

\subsubsection{Curriculum and Multi-fidelity Objective}
We construct our multi-fidelity function using a time-based curriculum proposed in \cite{peng2018deepmimic,pytorchrl}.
Given the normalized task difficulty scalar $z$, task difficulty is given by maximum episode length $T(z) = 30 + 570 z$, which ranges from $1$ to $20$ seconds. Our multi-fidelity objective considers both policy performance and sample efficiency, which is defined as:
\begin{equation}
    f(x, z) = \frac{\sum_{t= 1}^{t=T(z)}r(s_{t},a_{t})}{T(z)}e^{-(\frac{N_{sample}}{C})} .
\end{equation}
$N_{sample}$ is the total number of samples consumed during training. Training is terminated when normalized expected return reaches a threshold pre-defined in original DeepMimic, or the normalized expected return does not increase for $500$ iterations. $C$ is a scaling factor set to $10^{7}$ and $2\times10^{7}$ for walk and backflip respectively, based on the scale of usual samples required by DeepMimic on different tasks. Note that original DeepMimic has a fixed annealing-based schedule for maximum episode length during training. This schedule can be too slow on easy tasks. We switch to an adaptive schedule to avoid wasting samples. Starting from $1$ second, episode length is increased by $0.5$ seconds every iteration until it reaches $T(z)$, as long as the current fail rate is below $20\%$. 

\begin{figure}
     \centering
     \begin{subfigure}[]{0.4\linewidth}
         \centering
         \includegraphics[width=\linewidth]{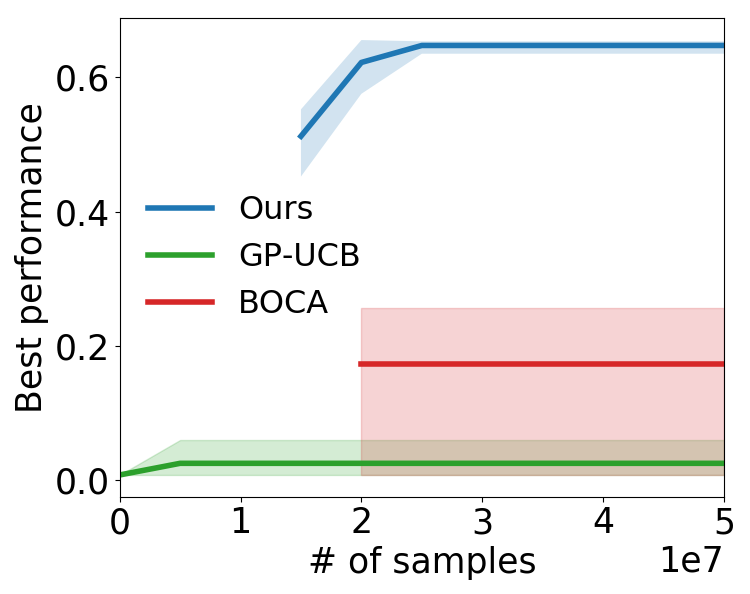}
         \caption{Walk}
         \label{fig:lr_comparision_walk}
     \end{subfigure}
     \begin{subfigure}[]{0.4\linewidth}
         \centering
         \includegraphics[width=\linewidth]{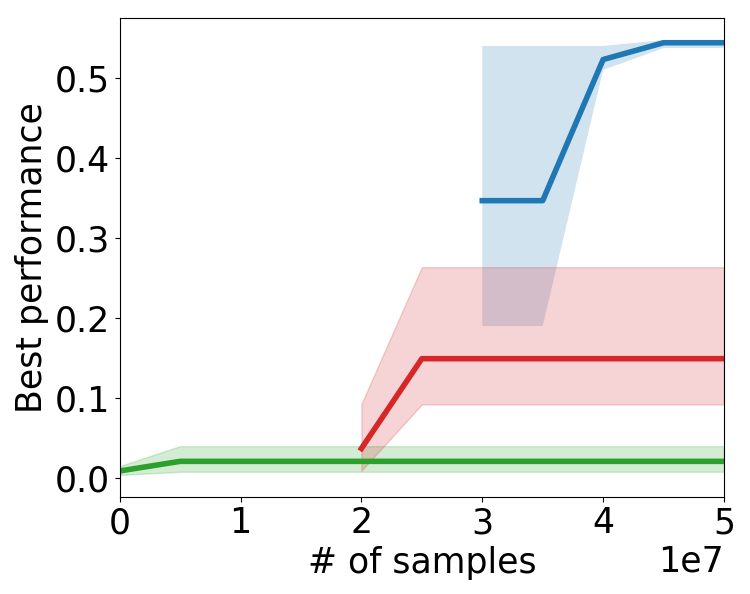}
         \caption{Backflip}
         \label{fig:lr_comparision_backflip}
     \end{subfigure}
         \caption{DeepMimic hyperparameter optimization results. We compare best performance over the number of training samples among CMFBO, GP-UCB and BOCA.}
        \label{fig:lr_comparison_deepmimic}
\end{figure}
\begin{figure}
     \centering
     \begin{subfigure}[]{0.4\linewidth}
         \centering
         \includegraphics[width=\linewidth]{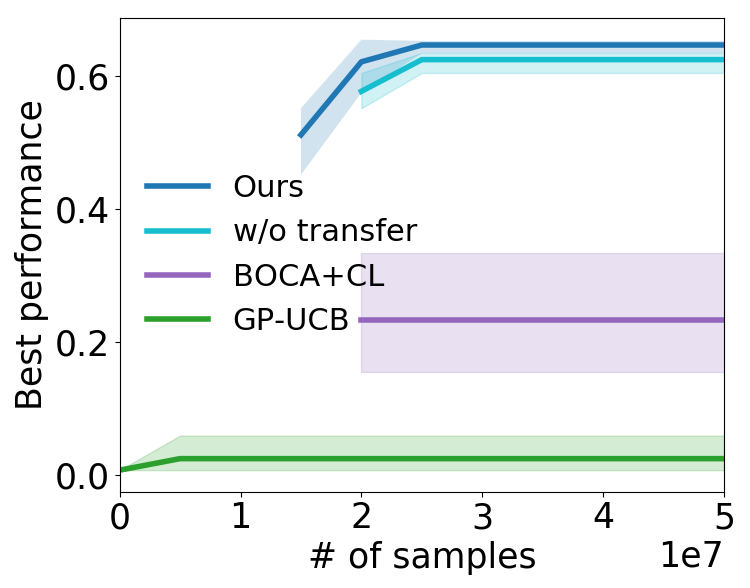}
         \caption{Walk}
         \label{fig:ablation_walk}
     \end{subfigure}
     \begin{subfigure}[]{0.4\linewidth}
         \centering
         \includegraphics[width=\linewidth]{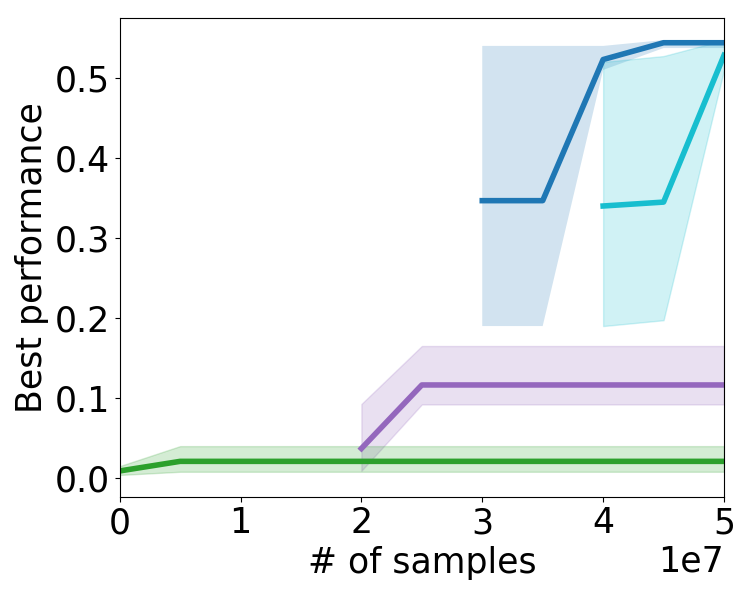}
         \caption{Backflip}
         \label{fig:abaltion_backflip}
     \end{subfigure}
         \caption{Ablation studies on DeepMimic hyperparameter optimization.}
        \label{fig:lr_ablation_deepmimic}
\end{figure}

\subsubsection{Comparison}
We compare our method against GP-UCB and BOCA in our experiments. For BOCA based on number of iterations as its fidelity criterion, maximum iterations on each fidelity $N(z)$ is set to $500 + 6500z$ and $500 + 11500z$ respectively for walk and backflip. We also show comparison with BOCA using our curriculum-based fidelity in our ablation study. Maximum allowed simulation steps $N_{total}$ is set to $5\times10^{7}$ for all methods. The optimization process takes about $16$ hours for all methods. 


\begin{table}[]
\begin{tabular}{ccc}
\hline
Parameters                      & Walk      & Backflip \\ \hline
Batch size                       & $172$       & $48$       \\
Policy updates per iteration      & $8$         & $10$       \\
Learning rate of actor network  & $1.68\times10^{-5}$ & $4.6\times10^{-5}$ \\
Learning rate of critic network & $2\times10^{-3}$    & $0.01$     \\
Weight decay                    & $5\times10^{-4}$    & $5\times10^{-4}$   \\
PPO clip rate                   & $0.18$      & $0.02$     \\
Maximum of the gradient norm    & $47.36$     & $12.30$    \\ \hline
\end{tabular}
\caption{Optimized hyperparameters for walk and backflip.}
\label{table::paras_walkBackflip}
\end{table}

\subsubsection{Results}
\begin{figure}
     \centering
     \begin{subfigure}[b]{0.4\linewidth}
         \centering
         \includegraphics[width=\linewidth]{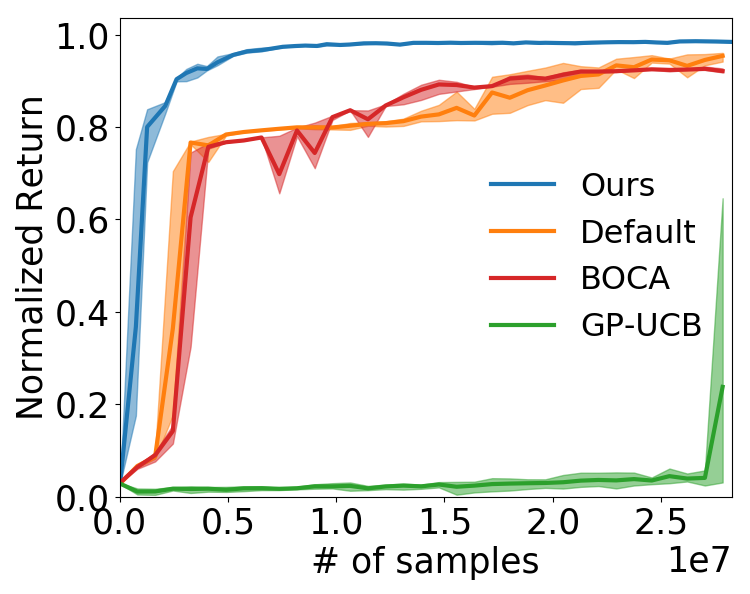}
         \caption{Walk}
         \label{fig:lr_walk}
     \end{subfigure}
     \begin{subfigure}[b]{0.4\linewidth}
         \centering
         \includegraphics[width=\linewidth]{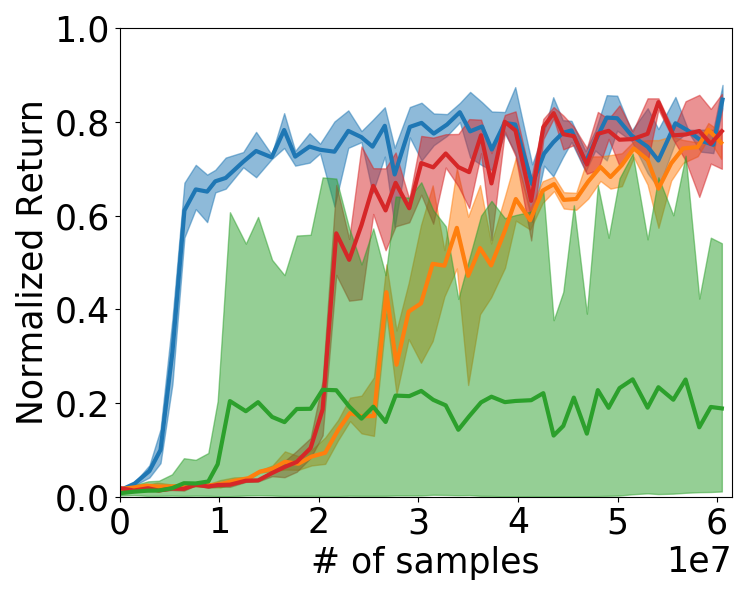}
         \caption{Backflip}
         \label{fig:lr_backflip}
     \end{subfigure}
         \caption{Learning curves of policies trained with hyperparameters from CMFBO, DeepMimic default settings, BOCA and GP-UCB.}
        \label{fig:comparison_deepmimic}
\end{figure}

As shown in figure \ref{fig:lr_comparison_deepmimic}, our method consistently outperforms other methods. CMFBO find different hyperparameters with different random seeds, but the performance is consistently good. Table \ref{table::paras_walkBackflip} shows the best optimized hyperparameters, which are quite different from the default ones listed in table \ref{table::paras_deepmimic}.
We select best hyperparameters optimized from each method and test their policy learning efficiency based on $3$ individual runs. Results are shown in figure \ref{fig:comparison_deepmimic}. 
GP-UCB fails to find hyperparameters with which policies could be learned successfully. Since DeepMimic is computationally expensive, only few evaluations can be performed in such single fidelity methods. BOCA performs better than GP-UCB and find hyperparameters as good as the default ones for walk task. It also finds hyperparameters converging $1.7$ times faster than default ones for backflip. 
Comparing to GP-UCB and BOCA, our method finds high-performing hyperparameters highly efficiently. Hyperparameters optimized by CMFBO requires $5-6$ times less simulation steps than default settings for successful policy learning. We note that by default DeepMimic takes about $8$ and $16$ hours respectively to train control policies until convergence for walk and backflip, while our method can find high-performing hyperparameters within $8$ and $13$ hours respectively for these tasks, even before the training of default DeepMimic finishes. We conduct the same set of ablation studies as in morphology optimization shown in figure \ref{fig:lr_ablation_deepmimic}, where we compare CMFBO against BOCA with curriculum-based fidelity and CMFBO without policy transfer. As expected, each component is essential to achieve the impressive efficiency of CMFBO. 

\begin{figure}
     \centering
     \begin{subfigure}[b]{0.4\linewidth}
         \centering
         \includegraphics[width=\linewidth]{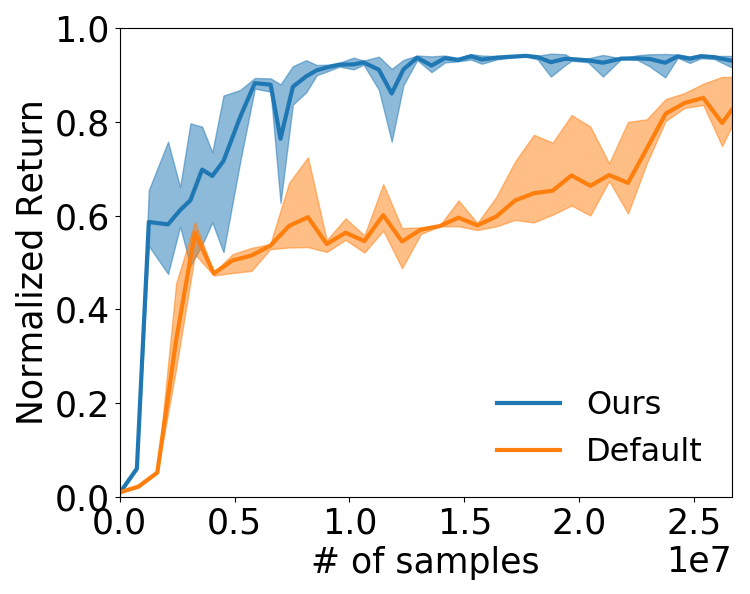}
         \caption{Run}
         \label{fig:lr_run}
     \end{subfigure}
     \begin{subfigure}[b]{0.4\linewidth}
         \centering
         \includegraphics[width=\linewidth]{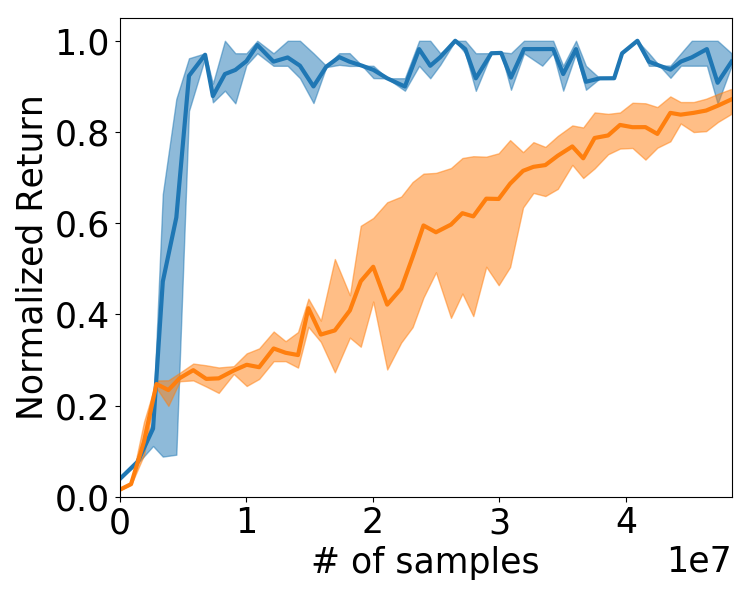}
         \caption{Cartwheel}
         \label{fig:lr_cartwheel}
     \end{subfigure}
         \caption{Learning curves of policy learning for other tasks with hyperparameters optimized by CMFBO on walk task comparing with DeepMimic default settings.}
        \label{fig::transfer_deepmimic}
\end{figure}

We show that hyperparameters optimized by CMFBO generalize to other tasks as well. Figure \ref{fig::transfer_deepmimic} 
illustrates the large performance gain of policy learning on run and cartwheel tasks using hyperparameters optimized by CMFBO on walk comparing with DeepMimic default settings.

CMFBO can also give us insights on the sensitivity of each hyperparameter. After optimization, the surrogate model shows small length scales along dimensions corresponding to policy updates per iteration, learning rate of critic network and weight decay. This implies that the performance is more sensitive to these hyperparameters within the given ranges.

We briefly analyze the computational cost distribution of CMFBO for DeepMimic hyperparameter optimization. The optimization of acquisition function usually finishes within three minutes, which is negligible comparing with the time required for DRL training. We summarize the number of training samples consumed on different fidelity levels in table \ref{table::distribution_walkBackflip}, which indicates that most computational resources are allocated to low fidelity approximations. Easy tasks require less time to train than difficult ones. For instance, in walk task of DeepMimic, training at the easiest and the original difficulty level with our optimized hyperparameters take roughly $20$ minutes and $80$ minutes respectively.



\begin{table}[]
\begin{tabular}{@{}lcc@{}}
\hline
                                                  & \multicolumn{1}{l}{Walk} & \multicolumn{1}{l}{Backflip} \\ \midrule
\multicolumn{1}{c}{\# of sampled hyperparameters} & $20$                       & $23$                           \\
\# of training samples on original tasks          & $6.1\times10^{6}$                      & $1.23\times10^{7}$                         \\
\# of training samples on low fidelities          & $4.39\times10^{7}$                     & $3.77\times10^{7}$                                                            \\ \hline
\end{tabular}
\caption{The distribution of training samples in CMFBO}
\label{table::distribution_walkBackflip}
\end{table}

\section{CONCLUSIONS}
This paper introduces the hyperparameter optimization problem in physics-based character animation. We present CMFBO, a novel multi-fidelity Bayesian optimization framework to achieve efficient optimization. Motivated by curriculum learning, we propose the use of task difficulty as an effective fidelity criterion, which enables relative performance of different hyperparameters to be accurately estimated even with low-fidelity cheap approximations. We enable efficient search space pruning through a progressive acquisition function focusing only on optimal regions. Transfer learning is further adopted to reduce evaluation cost of function queries. Through extensive experiments, we demonstrate that CMFBO is more efficient than state-of-the-art hyperparameter optimization algorithms. In particular, we show that hyperparameters optimized through CMFBO result in at least 5x performance gain comparing to original author-released settings in DeepMimic. We believe that our method could serve as stepping-stone for automatic tuning of physics-based character animation systems and free researchers or engineers from laborious and exhausting tuning works.

There are several interesting future directions worth exploring. First, the search space grows exponentially with the dimension of hyperparameters. We could incorporate some recent machine learning and optimization techniques, like random embedding \cite{wang2013bayesian}, to scale our method to even higher dimensions requiring much larger computation budgets, such as optimizing the complex routing of muscles for full-body characters. Second, our current method is limited to Euclidean search space. It would be interesting to extend our method to general Non-Euclidean search space such as Riemannian manifolds in \cite{jaquier2020bayesian}. Last but not least, although our method is demonstrated on simulated characters, it could possibly be extended to real-world automatic robot design with the recent success of 'sim to real' transfer learning \cite{tan2018sim,yu2019sim}.


\bibliographystyle{ACM-Reference-Format}
\bibliography{CMFBO}

\end{document}